\documentclass{icrc29}
\usepackage{times,here,amssymb,amsmath}
\usepackage{graphicx,lineno} 
\begin{document}
\title[Uncertainties in event reconstruction and $S(1000)$ determination 
  by the Pierre Auger Observatory]
  {Statistical and systematic uncertainties in the event
  reconstruction and {\em S}(1000) determination by the Pierre Auger
  surface detector}
\author[Pierre Auger Collaboration] {Pierre Auger Collaboration\\
Pierre Auger Observatory, Av. San Mart{\'\i}n Norte 304, (5613), Malarg\"ue,
    Argentina
}
\presenter{Presenter: Piera L. Ghia (piera.ghia@lngs.infn.it) \
ita-ghia-P-abs1-he14-oral}

\maketitle

\begin{abstract}
We discuss the statistical and systematic uncertainties in the event
reconstruction (core location, and determination of $S(1000)$, i.e., the
signal at a distance of 1000 m from the shower core) by the
Pierre Auger surface detector for showers with zenith angle less than 60
degrees. The method is based on a maximum likelihood method
where the reference lateral distribution function is obtained through the experimental data.
We also discuss $S(1000)$ as primary energy estimator. 
\end{abstract}

\section{Introduction}
The Pierre Auger Observatory, combining fluorescence telescopes with 
an extensive air shower array of water
Cherenkov detectors, is well apt to the study 
of cosmic rays at the highest energies ($E>10^{18}$ eV).
The use of the two techniques allows the determination of 
the primary energy improved with respect to former arrays
and with no basic dependence on interaction models and hypothesis on the
primary composition. The technique is based on the calibration of the energy
estimator of the surface detector (SD) through the fluorescence detector
data (FD): the SD provides the required statistics thanks to its collecting area and
its much larger duty cycle (around 100\%). The signal measured at a specific
distance from the shower axis, $S(r)$, is well established as an energy estimator
for the surface detector: in the Auger observatory $r$ is 1000 m.
The accuracy in the determination of such estimator depends on the detector
resolution and sampling fluctuations,
and on shower fluctuations in the longitudinal development,
since the measurement is performed at a single, fixed atmospheric depth. Moreover,
the SD energy estimator is not directly measured, but interpolated by a fit
to a lateral distribution function (hereafter called LDF).
In this paper, we present the results of a  
study of the accuracies in the event reconstruction and in the
determination of the energy estimator by the SD, in view 
of the measurement of the energy spectrum. With respect to the uncertainties
arising from the detector sampling different kinds of shower particles 
and from the LDF shape,
since it is not possible to rely on theoretical predictions, these studies are
performed using the experimental data themselves.
On the other side, simulations are needed to evaluate
the uncertainties due to cascade fluctuations and to justify the choice of
$S(1000)$ as energy estimator.\\
After giving in section 2 a brief description of the Pierre Auger array,
we will present in section 3 the event reconstruction technique and the
systematic and statistical accuracies in the determination of $S(1000)$ and
core position. We will discuss in section 4 the effects of shower
fluctuations and the stability of $S(1000)$ with respect to such
fluctuations.

\section{The Pierre Auger surface detector}
The surface detector of the Pierre Auger Observatory is extensively
described in \cite{ber05}. 
We will limit here to describe the characteristics of the detector
relevant to the analysis discussed in the following sections.\\
The water in each Cherenkov detector is viewed with three 9''
diameter photomultipliers. Signals are extracted from the anode and from
the last dynode (amplified by a factor 40, resulting in a dynode-to-anode
ratio of 32) and are then converted, after
appropriate calibration (see \cite{all05} for a
detailed description), in units of vertical equivalent muons (VEM).
The VEM is the unit used
in the subsequent analysis, both for the LDF and the reconstruction of
$S(1000)$.\\
A hierarchical trigger sequence (fully described in \cite{len05}) is
used to identify a cosmic ray event. The lowest level is imposed on each
single station, which is fired if the signal satisfies either 
a 3-fold coincidence of 1.75 VEM on each PMT
dynode or a 2-fold coincidence of 13 bins exceeding 0.2 VEM within a 3$\mu$s
window. 
The highest level triggers are
realized to select physical events, and, for each of them, the stations
($\ge 3$) which are not due to chance coincidences. 
The event reconstruction is
applied to all the physical events, and among them ``quality''
events are selected.
The ``quality'' cuts, which will be used in the following analysis, 
select showers 
with the highest signal recorded in a tank surrounded by at least 
five operating ones, and the reconstructed core within a triangle of
working stations.\\
The arrival directions are measured from the relative arrival times of the
signals in the selected stations \cite{bon05}, while
the core position, ($x_{\textrm c}, y_{\textrm c}$), 
and $S(1000)$ are determined from the water Cherenkov
signals recorded by each selected station, through a fit 
to a formula describing the LDF.

\section{Event reconstruction: core position and {\em S}(1000)}
 
\subsection{The reconstruction technique}
The reconstruction is based on a maximum likelihood method, in which the
signals measured by each station, $S_{\textrm {mea}}$, are compared with those
expected from a
lateral distribution function (LDF), $S$. Determination of the LDF is described
in \cite{bau05} and it is represented by a
Nishimura-Kamata-Greisen form:\\
$ S(r)=A \left [ \frac{r}{r_{\textrm s}} \left (1+\frac{r}{r_{\textrm s}}\right ) \right ] ^{-\beta}$,
with $r_{\textrm s}=700$ m, $\beta=2.4-0.9\cdot (\textrm{sec} \theta-1)$ and
$A=S(1000)\cdot 3.47^{\beta}$.\\
The likelihood function includes:\\
(i) a term due to fired stations.
This is is given by:
$\displaystyle\sum_{\textrm{fired}}\frac{{(S_{\textrm{mea}}-S)}^2}{\sigma^2_{\textrm{fired}}}$
where $\sigma_{\textrm{fired}}=1.06 \sqrt{S}$ is the uncertainty in the signal
measurement 
\cite{tok03}. For fired stations with a saturated anode signal
in at least one channel, $S_{\textrm{mea}}$ is the
signal deduced using the undershoots on the dynode 
and anode channels, and due to the 
coupling capacitors of the bases. In this case
$\sigma_{\textrm{fired}}=0.08 S$ and $\sigma_{\textrm{fired}}=0.13 S$, for anode
undershoot $U_{\textrm a}\le 1$ ADC channel and $U_{\textrm a}>1$ ADC 
channel, respectively.\\[0.1cm]
(ii) a term due to the stations which
are below trigger threshold, up to 10 km from a triggered
one. 
The contribution of these stations is
described by a Poisson law, with no-detection probability, $P$, up to 4 VEM:
$-2\textrm{ln}(P)=2S-2\textrm{ln}(1+S+S^2/2!+S^3/3!+S^4/4!)$
\subsection{Uncertainties in the determination of the core position and
  {\em S}(1000)}
\begin{figure}[H]
\begin{minipage}{0.49\linewidth}
\begin{center}
\vspace{-1.cm}
\includegraphics[width=8.3cm]{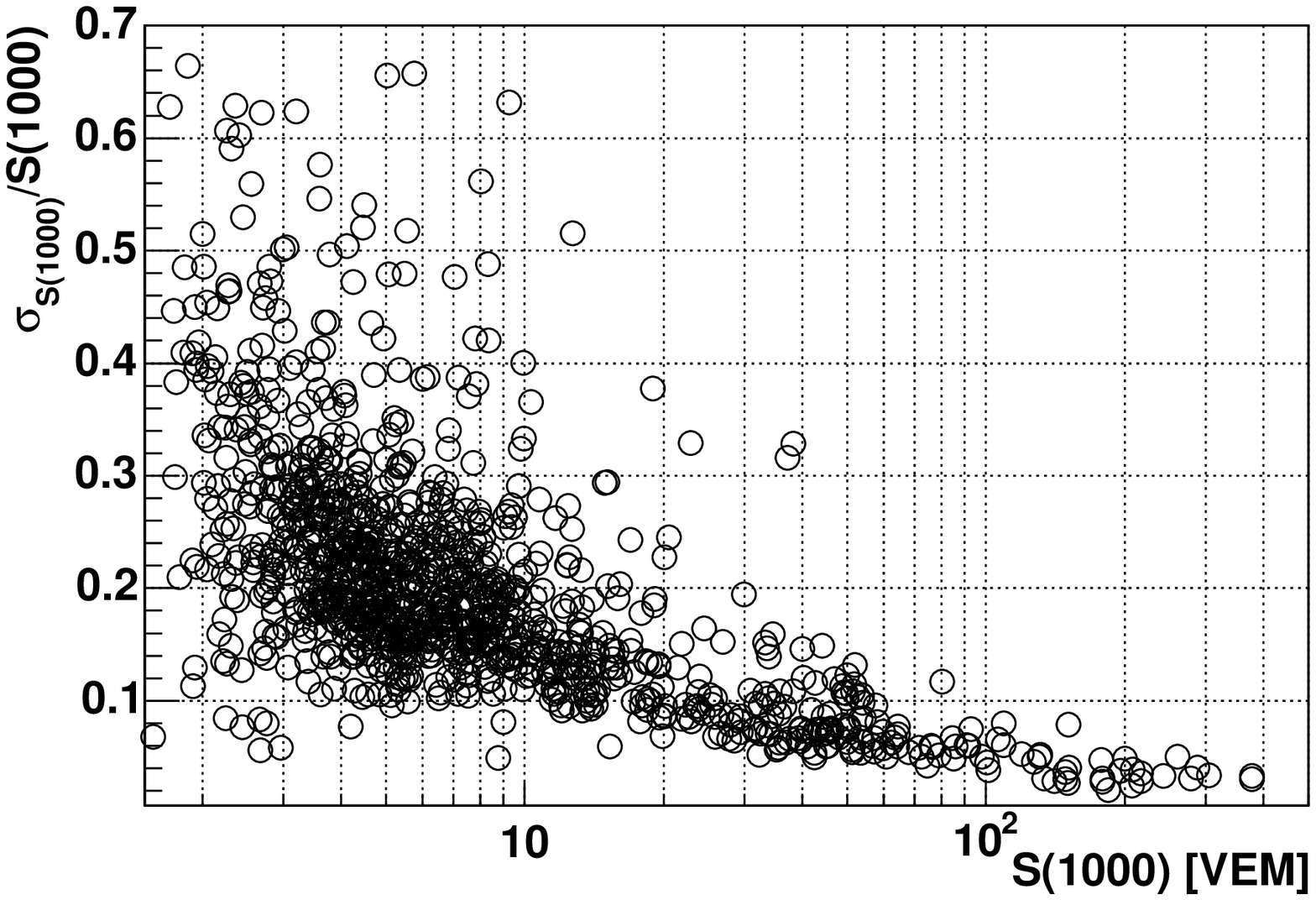}
\vspace{-0.9cm}
\caption{\label{fig1} $\sigma_{S\textrm{(1000)}}$/$S(1000)$ vs $S(1000)$, 
obtained by the ``fluctuation'' method (see text).}
\end{center}
\end{minipage}
\hspace{0.2cm}
\begin{minipage}{0.49\linewidth}
\begin{center}
\vspace{-0.5cm}
\includegraphics[width=8.4cm]{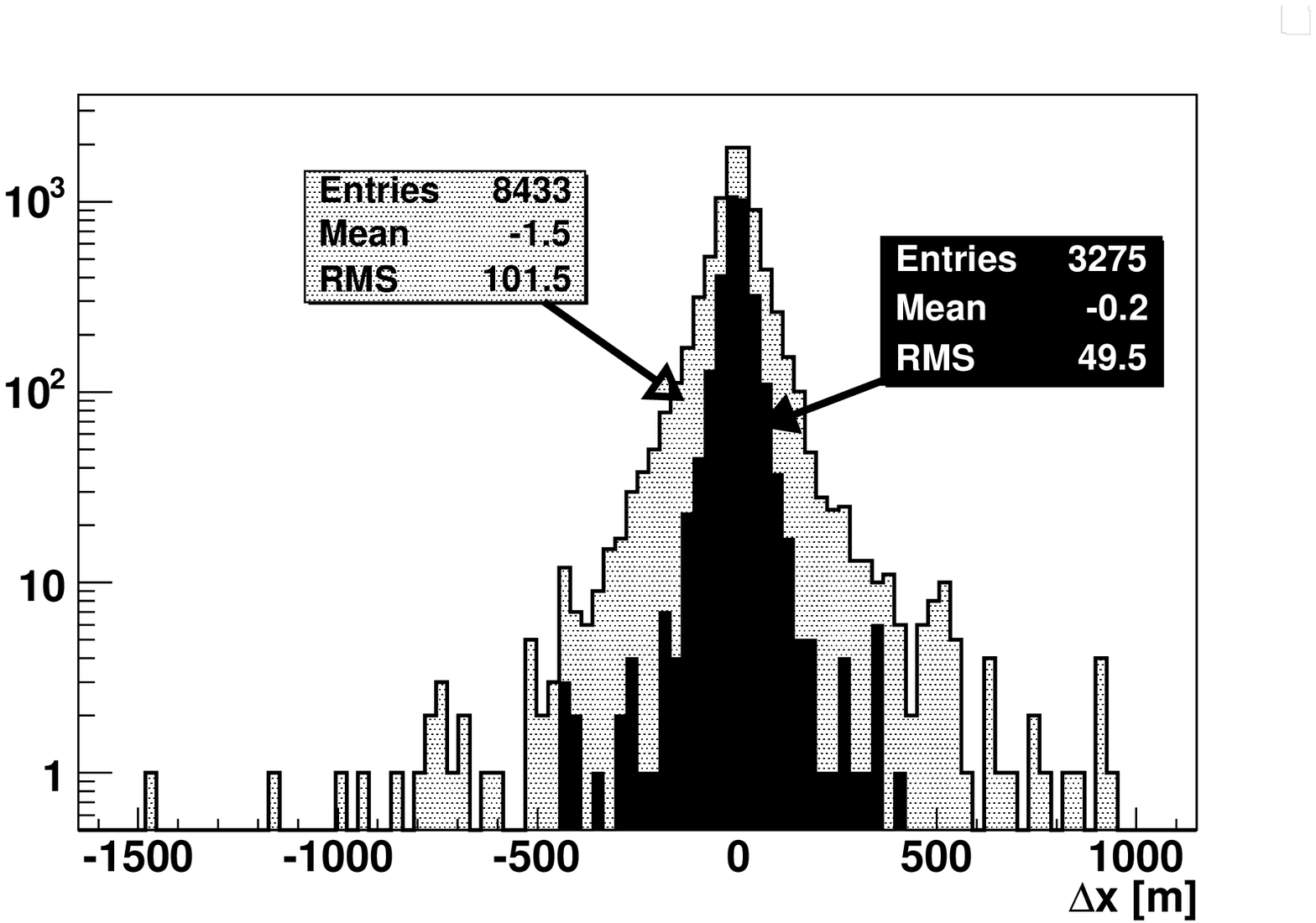}
\vspace{-0.9cm}
\caption{\label{fig2}Distribution of differences in core 
locations $\Delta x$: the gray
histogram comprises all events, while the black one those with $S(1000)>$ 30.}
\end{center}
\end{minipage}
\end{figure}

The statistical accuracies in the measurement of 
($x_{\textrm c},y_{\textrm c})$ and $S(1000)$ are
obtained by fluctuating the signals detected in each
station, $S_{\textrm {mea}}$,  adopting a Gaussian law with mean
value corresponding to $S_{\textrm {mea}}$ and
r.m.s. $\sigma=\sigma_{\textrm {fired}}$. The
reconstruction procedure is then applied to the ``fluctuated'' event, and
repeated 30 times. The widths of the distributions of 
$S(1000)$, $x_{\textrm c}$ and
$y_{\textrm c}$ provide the statistical uncertainty: we show in figure \ref{fig1} the
behavior of
$\sigma_{S\textrm{(1000)}}/S(1000)$ vs $S(1000)$, and in figure \ref{fig2} the
distribution of the differences in core location.\\
The dependence of systematic uncertainties on the determination of ($x_{\textrm c},
y_{\textrm c}$) and $S(1000)$ arising from the assumed form of the LDF has been
investigated. 
The systematic differences are at the level of about 4\% with
fluctuations around 4\%
for either a log-log parabola
or the NKG-like form with different $\beta$ \cite{bau05}, thus
showing that the determination of $S(1000)$ is quite stable
with respect to the choice of the LDF.
\hspace{0.5cm} 
\begin{figure}[H]
\begin{minipage}{0.48\linewidth}
\vspace{-0.8cm}
We finally studied the impact of  the ``quality cuts'' on the event
reconstruction. To this purpose, we
considered only events with the maximum signal detected by a tank surrounded by
a full circle of operating detectors, i.e., six. We hence artificially
switched off one or more of the fired stations
to simulate a shift of the event toward the boundaries, 
or to simulate the effect of a missing internal station.
In any case we preserved the quality criterion of five operating tanks
surrounding the one with the maximum signal, and 
we repeated the reconstruction procedure.
We show in figure 
\ref{fig4} the ratio between 
$S(1000)$ after the artificial boundary shift or loss of an internal tank, and the
original $S(1000)$:
the systematic difference is $\approx$ 2\% 
while the fluctuations are $\approx$ 8\%.
\end{minipage}
\hspace{0.2cm}
\begin{minipage}{0.5\linewidth}
\begin{center}
\vspace{-1.2cm}
\includegraphics[width=8.5cm]{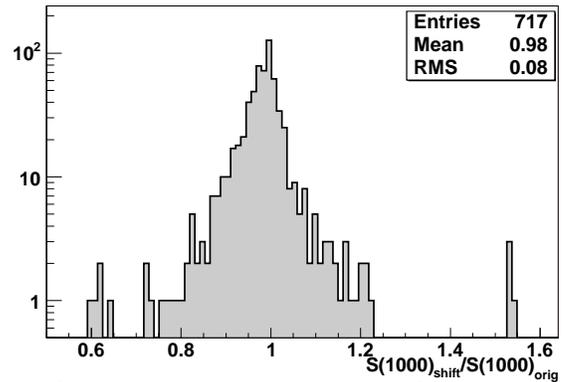}
\vspace{-1cm}
\caption{\label{fig4} Distribution of the ratio between $S(1000)$\
  reconstructed after the
  artificial shift of the event or artificial loss of an internal tank and
  the original $S(1000)$.}
\end{center}
\end{minipage}
\end{figure}
\vspace{-0.5cm}
This shows that
the ``quality cuts'' ensure a good uniformity of the reconstruction
accuracy with respect to the local position of the event inside the array,
or to the possible absence of an internal station.

\section{Shower-by-shower fluctuations in {\em S}(1000)}
\begin{figure}[H]
\begin{minipage}{0.49\linewidth}
\begin{center}
\vspace{-0.5cm}
\includegraphics[width=8.0cm]{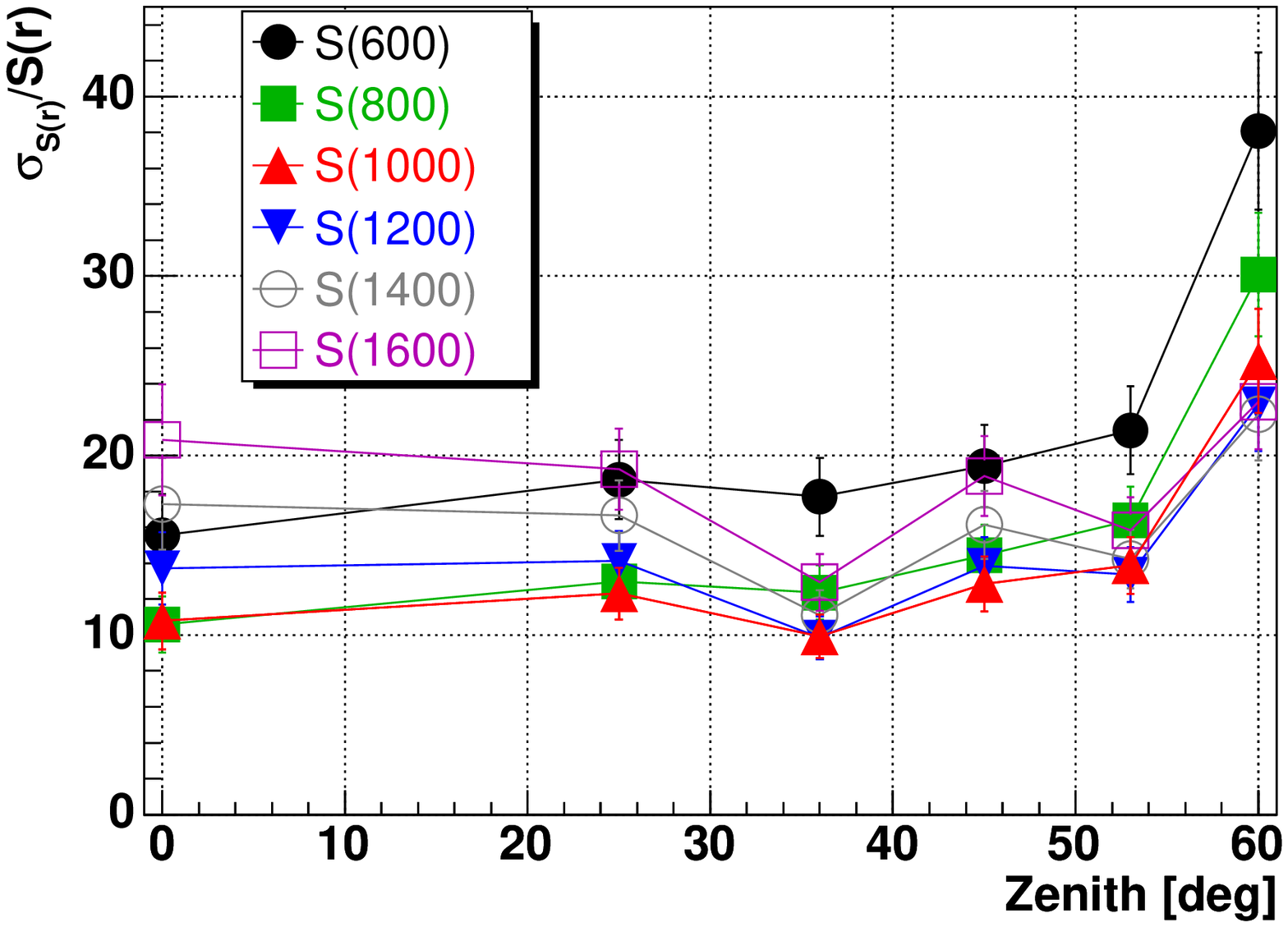}
\vspace{-0.8cm}
\caption{\label{fig5}Statistical uncertainty in $S(r)$, for various values of
  $r$, vs zenith angle, for 10 EeV primary protons (QGSJET).}
\end{center}
\end{minipage}
\hspace{0.3cm}
\begin{minipage}{0.48\linewidth}
\begin{center}
\vspace{-0.3cm}
\includegraphics[width=8.0cm]{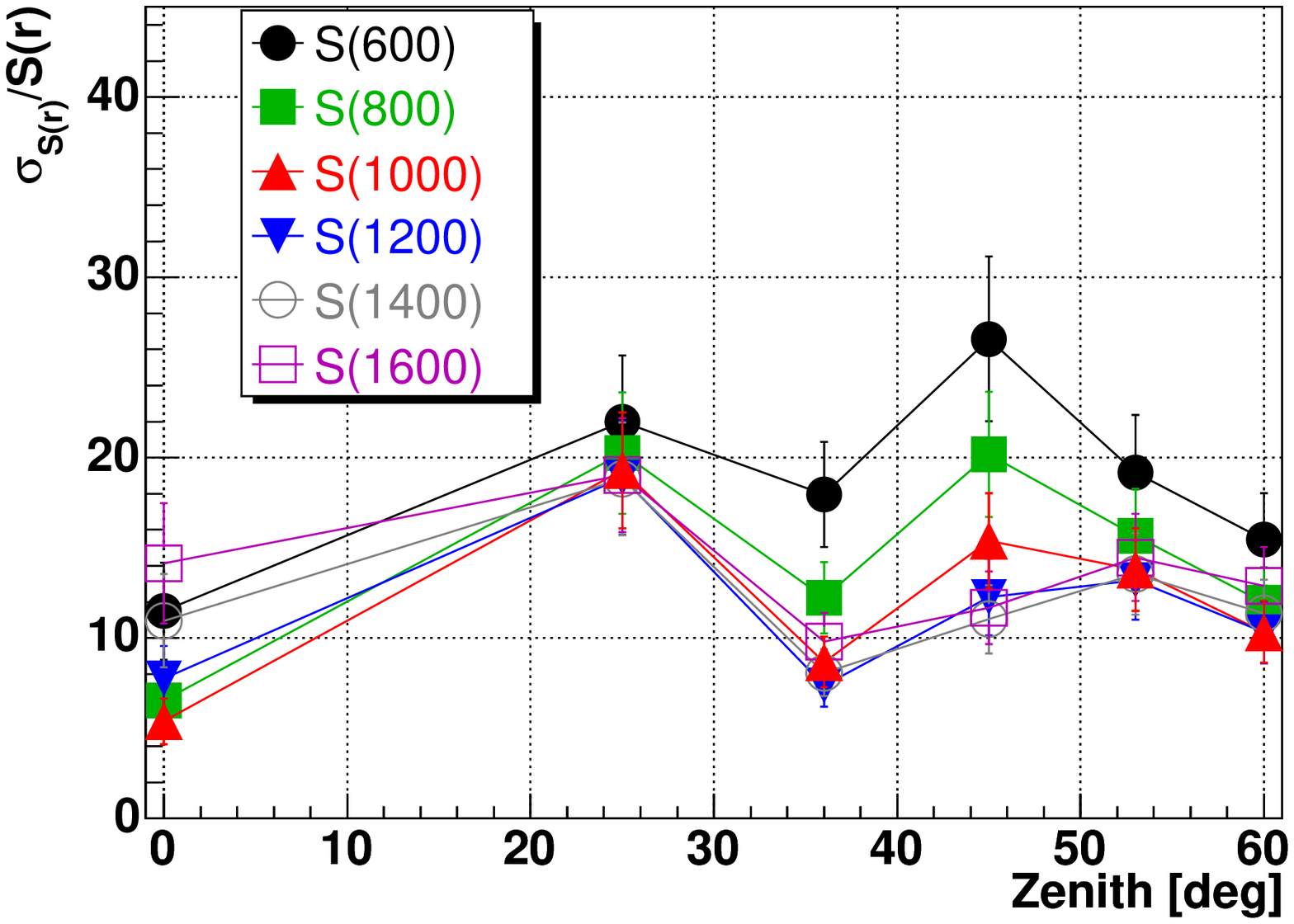}
\vspace{-0.8cm}
\caption{\label{fig6} Statistical uncertainty in $S(r)$, for various values of
  $r$, vs zenith angle, for 10 EeV primary irons (QGSJET).}
\end{center}
\end{minipage}
\end{figure}

Intrinsic fluctuations in the process of development of the showers in the
atmosphere are a further source of uncertainties in $S(1000)$, different for
different primary cosmic ray mass.
 Such fluctuations were studied by Monte
Carlo simulations using the AIRES package \cite{aires}. The air showers were simulated 
for various energies, zenith angles, primary types and interaction models 
(QGSJET and SIBYLL). The detector response was simulated using a 
lookup table derived from GEANT4 \cite{geant}. \\
Figures \ref{fig5} and \ref{fig6} show the r.m.s. of the $S(r)$ 
distributions, at 10 EeV primary energy, 
as a function of the zenith angle, for $r$= 600$\div$1600 m, for primary protons
and irons, respectively, and QGSJET as interaction model. 
These plots show that the shower-by-shower fluctuations are minimized for 
$S(1000)$, which is the main reason for choosing it as primary energy
estimator. 
\begin{figure}[H]
\begin{minipage}{0.51\linewidth}
\vspace{-0.4cm}
Figure \ref{fig7} shows the statistical uncertainty in $S(1000)$, due to
shower development, as a
function of $S(1000)$ for proton and iron primaries, at $\theta=36^\circ$,
using QGSJET as interaction model.

\section{Conclusions}
We have shown that for the geometry of the surface 
detector of the Pierre Auger Observatory
the water Cherenkov signal at a distance of 1000 m from
the shower core, $S(1000)$, is the best energy estimator, for all 
primaries and zenith angles.\\
With respect to the determination of
$S(1000)$, the statistical uncertainties at $S(1000)$ $\approx 30$ VEM
(corresponding to primary energy around $5\cdot 10^{18}$ eV) are of the order of
10\% (being around 50 m for the core location). 
The
\end{minipage}
\hspace{0.2cm}
\begin{minipage}{0.47\linewidth}
\begin{center}
\vspace{-1.0cm}
\includegraphics[width=7.8cm]{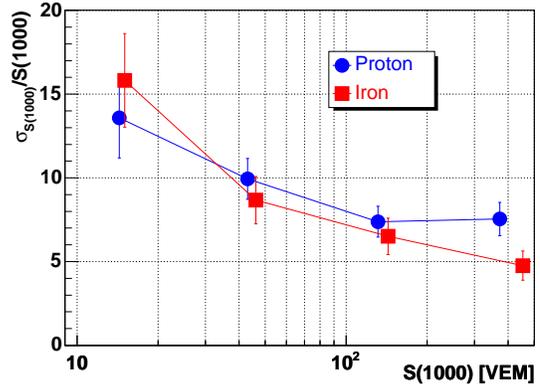}
\vspace{-0.8cm}
\caption{\label{fig7} Statistical uncertainty in $S(1000)$ vs $S(1000)$,
  for proton primaries (circles) and iron ones (squares) (QGSJET).}
\end{center}
\end{minipage}
\end{figure}
\vspace{-0.5cm}
systematic uncertainties due
to the assumed LDF form are $<4\%$, while those due to event 
sampling within the array
or to a missing internal tank 
give contributions, at most, of the order of the statistical
ones. 
Finally, the fluctuations of $S(1000)$ due to shower development and to event
reconstruction are of comparable order.


\begin{thebibliography}{99}

\bibitem{ber05}
Pierre Auger Collaboration, 29th ICRC, Pune (2005) arg-bertou-X-abs1-he14-oral

\bibitem{all05}
M. Aglietta {\it et al.}, Pierre Auger Collaboration, 
29th ICRC, Pune (2005) usa-allison-P-abs1-he14-poster

\bibitem{len05}
I. Lhenry-Yvon {\it et al.}, Pierre Auger Collaboration, 29th ICRC, Pune
(2005) usa-bauleo-PM-abs1-he14-poster

\bibitem{bon05}
Pierre Auger Collaboration, 29th ICRC, Pune
(2005) bra-bonifazi-C-abs1-he14-oral

\bibitem{bau05}
D. Barnhill {\it et al.}, Pierre Auger Collaboration, 29th ICRC, Pune
(2005) usa-bauleo-PM-abs2-he14-poster

\bibitem{tok03}
T. Yamamoto {\it et al.}, Pierre Auger Collaboration, Proceedings 28th ICRC
(Tsukuba, Japan), (2003) 469

\bibitem{aires}
S. Sciutto, AIRES User's Manual available
at http://www.fisica.unlp.edu.ar/auger/aires

\bibitem{geant}
S. Agostinelli {\it et al.}, GEANT4 Collaboration, Nucl. Instr. Meth A, 506
(2003) 250

\end{thebibliography}
\end{document}